Hellenic Complex Systems Laboratory

# A nonlinear component of the analytical error

Technical Report III

Aristides T. Hatjimihail

2001

## Abstract

**Background**

In clinical chemistry, a number of studies shows that the probability of very large errors is much greater than expected from the Gaussian distribution. In addition, it has been empirically found that the behavior of nonlinear complex systems is often asymptotically exponential. Consequently, we may assume that the error of some analytical systems may be approximated by the sum of a linear component of error with Gaussian distribution and a nonlinear component with Laplacian. Then, the probability density function (pdf) of the total error is approximated by the convolution integral of the Gaussian and the Laplacian pdf.

**Methods**

To explore the assumption of a nonlinear component of the analytical error I have evaluated this distribution and calculated various quality control related statistics with numerical methods.

**Results**

Large errors are much more probable with the proposed distribution than with the Gaussian. Simulated series of measurements with the proposed distribution often meet the criteria for normality. The critical errors and the probabilities for critical error detection are less than the respective ones of the Gaussian distribution. The probabilities for false rejection are greater.

**Conclusion**

To optimize the quality control planning process, we should explore the possibility that there exists a nonlinear component of the analytical error.

## 1. Introduction

In clinical chemistry, it is usually assumed that the distribution of the analytical error is normal or Gaussian. Therefore, very large analytical errors are considered very improbable. This assumption of the normality of the analytical error is implied by the successful application of the linearization to the analytical systems. Accordingly, the analytical error may be represented as the sum of many small contributions none of which contributes very much to the total error.

A number of studies show that the probability of very large analytical errors is much greater than expected. Witte et al. have shown that the average probability of errors greater than seven standard deviations (SD) equals $4.47 \cdot 10^{-4}$ [13]. This probability is $1.7 \cdot 10^{8}$ times greater than it is expected from the Gaussian distribution.

In a brief outline of the emerging field of the statistics of complexity, Goldenfeld and Kadanoff have pointed out that the behavior of nonlinear complex systems is rarely normal [3]. It has been empirically found that the probability density function (pdf) of these systems for large deviations takes the form of an exponential or Laplacian distribution. Large errors are much more probable with the exponential and Laplacian distributions than with the Gaussian.

Therefore, it is possible that there are nonlinear components of the analytical error of some analytical systems, contributing substantially to the total error. Consequently, we may assume that the error of some analytical systems may be approximated by the sum of a linear component of error with Gaussian distribution and a nonlinear component with Laplacian distribution. Assuming that these components have means equal to zero, and are statistically independent, then the pdf *f(z,l)* of the total error of these analytical systems is approximated by the convolution integral of the Gaussian pdf with standard deviation $s_g$, and the Laplacian pdf, with a scale parameter $l \cdot s_g$, where *l* is a weighting factor :

$$f(z,l) = \int_{-\infty}^{+\infty} \frac{1}{\sqrt{2\pi}\, s_g} e^{-\frac{1}{2}\left(\frac{x}{s_g}\right)^2} \frac{1}{2\, l s_g} e^{-\frac{|z-x|}{2\, l s_g}} dx$$

The variance and the mean of the distribution are respectively:

$$Var(l) = s_g^2 + 2\, l^2 s_g^2$$

$$\mu(l) = 0$$

The maximum likelihood function *ml(l)* given a series of *n* measurements $x_i$ with the proposed pdf $f(x_i,l)$, a weighting factor *l*, and a standard deviation *s*, is:

$$ml(l) = \sum_{i=1}^{n} Log(f(x_i,l)) =$$

$$\sum_{i=1}^{n} Log\left(\int_{-\infty}^{+\infty} \frac{1}{\sqrt{2\pi}\,\frac{s}{\sqrt{1+2l^2}}} e^{-\frac{1}{2}\left(\frac{x_i}{\frac{s}{\sqrt{1+2l^2}}}\right)^2} \frac{1}{2l\,\frac{s}{\sqrt{1+2l^2}}} e^{-\frac{|x_i-y|}{2l\frac{s}{\sqrt{1+2l^2}}}} dy\right)$$

This study presents this pdf and compares it with the Gaussian distribution. In addition, it presents plots of quality control (QC) related statistics for various values of *l*, including the probabilities for critical error detection. As critical errors are considered the maximum allowable random and systematic analytical errors, “defined in such a way that an upper bound has been set on the (clinical) type I error” [7]. The clinical type I error is the probability for rejection of the true hypothesis that there is no significant change of the analyte of a patient. The hypothesis is rejected when the observed change of the analyte is greater than the maximum medically allowable analytical error [5].

## 2. Methods

The proposed pdf and the respective cumulative distribution function (cdf) were evaluated using numerical methods. The critical errors were calculated as previously described [4], assuming that the upper bound of the probability of a type I clinical error was equal to 0.01. The probabilities for error detection and for false rejection were calculated with numerical methods, assuming one control measurement per run, and that the control measurements were independent and identically distributed.

As quality control rules $S(y)$ are defined these that reject an analytical run if the absolute value of the difference of at least one control measurement from the expected mean exceeds $y$ standard deviations (SD).

The normality hypothesis was tested using the measures of skewness and kurtosis, on the significance level of 0.01 [2]. The probabilities for rejection of this hypothesis were estimated using 10,000 simulated series of 100 or 1000 measurements respectively, with the proposed distribution.

The numerical computations, the simulations, and the plots were done with Mathematica®, Version 4.1 [14] (Wolfram Research, Champaign, Illinois, USA).

## 3. Results

The results are presented on Table 1 and on Figures 1-15. The errors are measured in SD units.

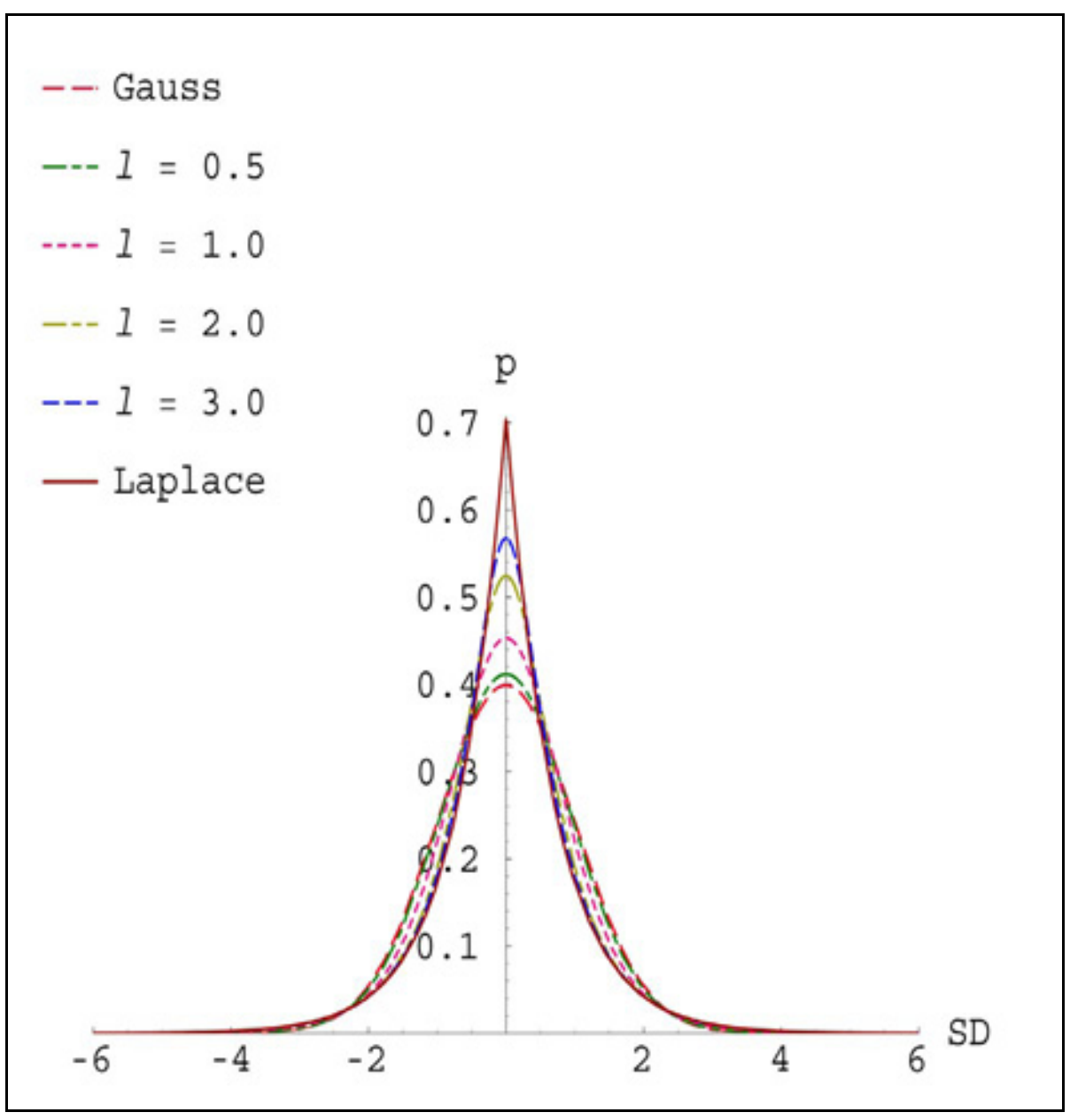

**Figure 1:** The Gaussian, the proposed, and the Laplacian pdfs

The pdfs of the Gaussian distribution, of the proposed distribution for $l$ equal to 0.5, 1.0, 2.0, and 3.0, and of the Laplacian distribution are presented on Figure 1, showing that the kurtosis of the proposed distribution is grater than the kurtosis of the Gaussian, and it is increasing with $l$.

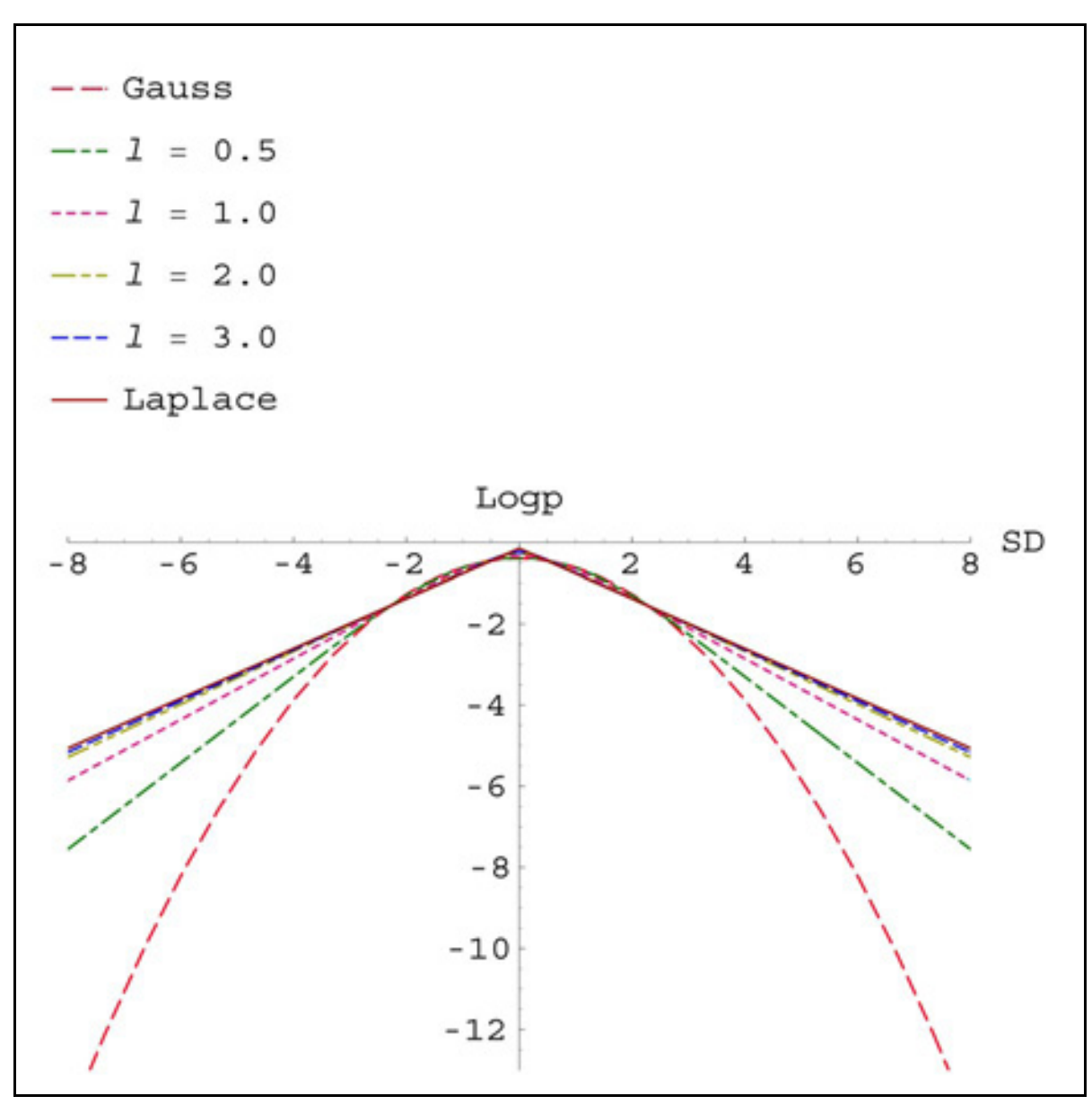

**Figure 2:** Logarithmic plots of the Gaussian, the proposed, and the Laplacian pdfs

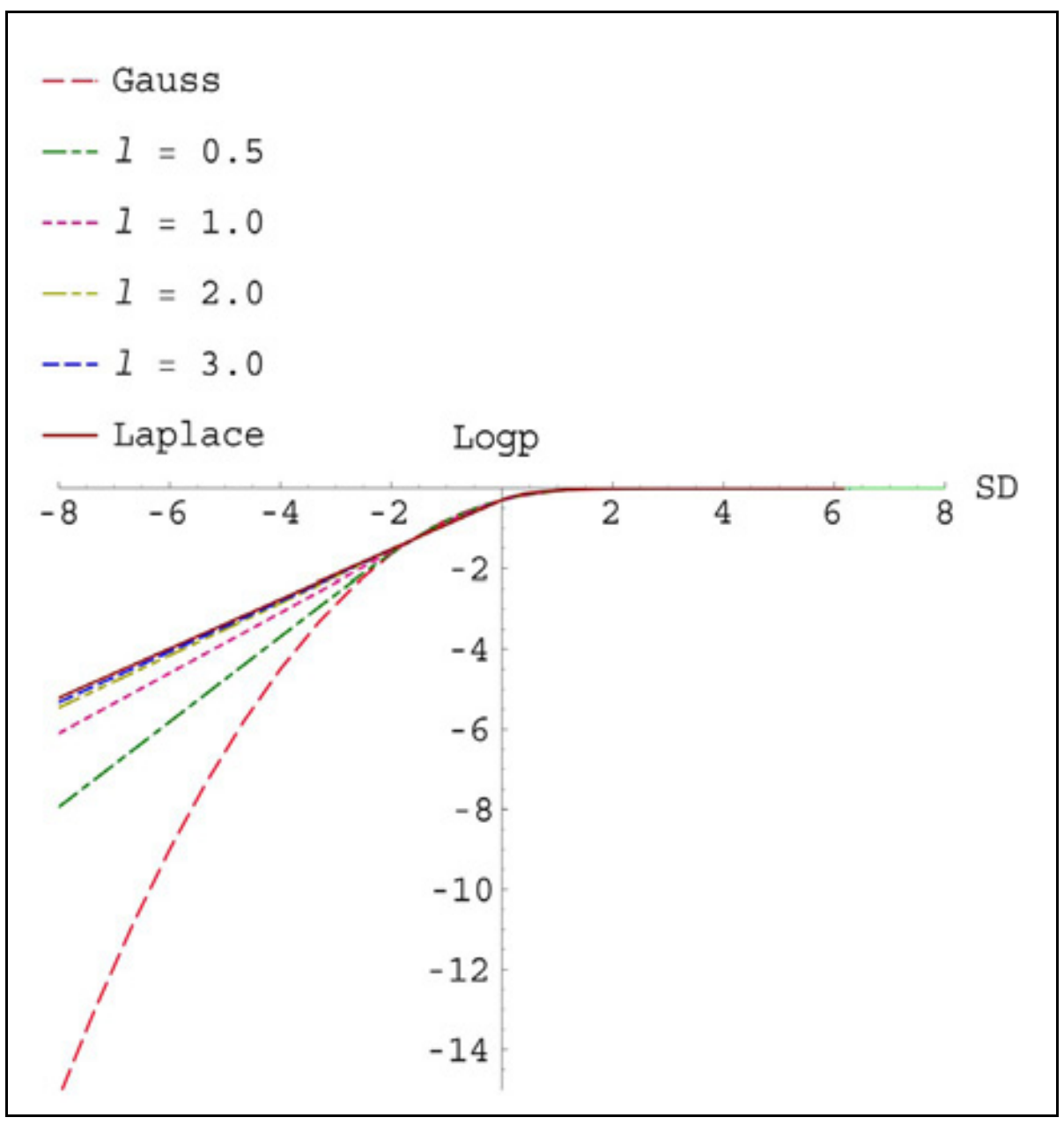

**Figure 3:** Logarithmic plots of the Gaussian, the proposed, and the Laplacian cdfs

Figures 2 and 3 are the logarithmic plots of the same pdfs and the respective cdfs, showing differences of some orders of magnitude of the probabilities of the very large errors.

The Figure 4 compares the Gaussian and the proposed distribution and shows the logarithmic plot of the ratio $r$ between the probabilities of an error $> x$ SD with the proposed distribution and an error $> x$ SD with the Gaussian distribution. Table 1 presents some of these ratios. They are increasing with $l$ and $x$.

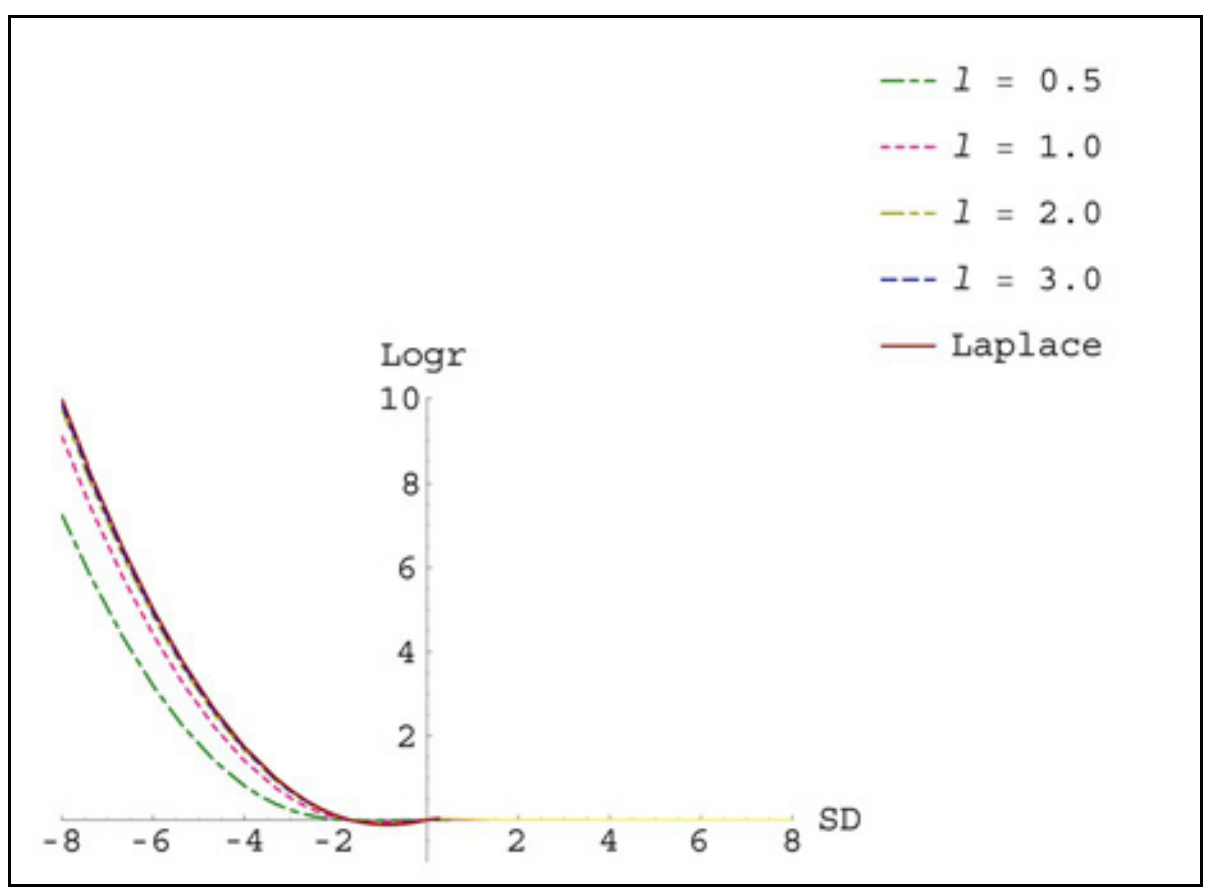

**Figure 4:** Logarithmic plots of the ratios *r* between the probability of an error >*x* SD with the proposed distribution and an error >*x* SD with the Gaussian distribution

**Table 1:** The table shows how much more probable is an error with the proposed distribution than with the Gaussian distribution

| | >2SD | >3SD | >4SD | > 5 SD | > 6 SD | >7 SD |
|---|---|---|---|---|---|---|
| $l$= 0.5 | 1.04 | 1.74 | 6.48 | 62 | $1.6\ 10^3$ | $1.0\ 10^5$ |
| $l$= 0.7 | 1.07 | 2.47 | 14.1 | $2.1\ 10^2$ | $8.1\ 10^3$ | $8.4\ 10^5$ |
| $l$= 1.0 | 1.13 | 3.38 | 25.5 | $5.0\ 10^2$ | $2.6\ 10^4$ | $3.5\ 10^6$ |
| $l$= 2.0 | 1.24 | 4.66 | 44.3 | $1.1\ 10^3$ | $7.1\ 10^4$ | $1.2\ 10^7$ |
| $l$= 3.0 | 1.27 | 5.01 | 49.9 | $1.3\ 10^3$ | $8.8\ 10^4$ | $1.6\ 10^7$ |

The Figure 5 presents the probabilities for rejection of the normality hypothesis with the proposed distribution versus *l*. The series are Gaussian for *l* = 0.

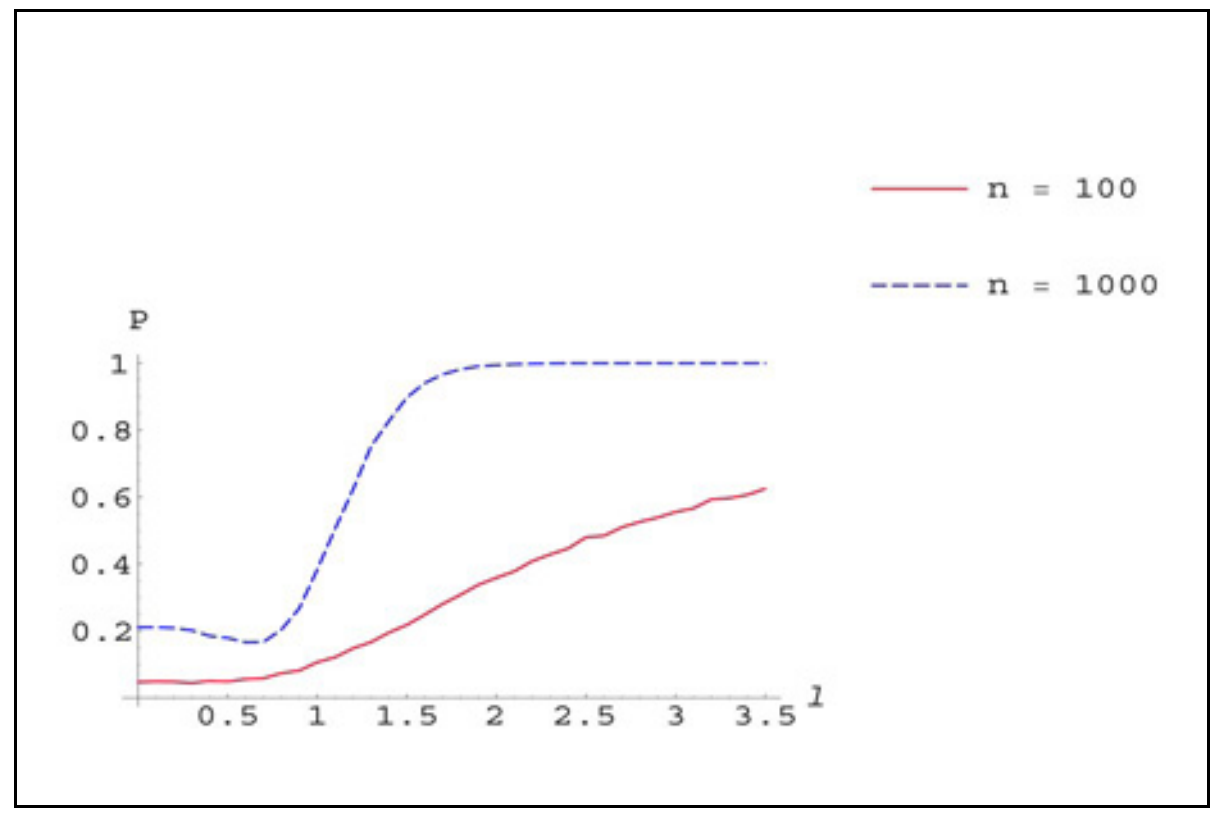

**Figure 5:** The estimated probabilities for rejection of the normality hypothesis of simulated series of 100 and 1000 measurements with the proposed distribution, versus *l*

The Figures 6-7 present the critical random and systematic errors with the Gaussian and the proposed distributions versus the total allowable analytical error. The critical errors are decreasing with *l*.

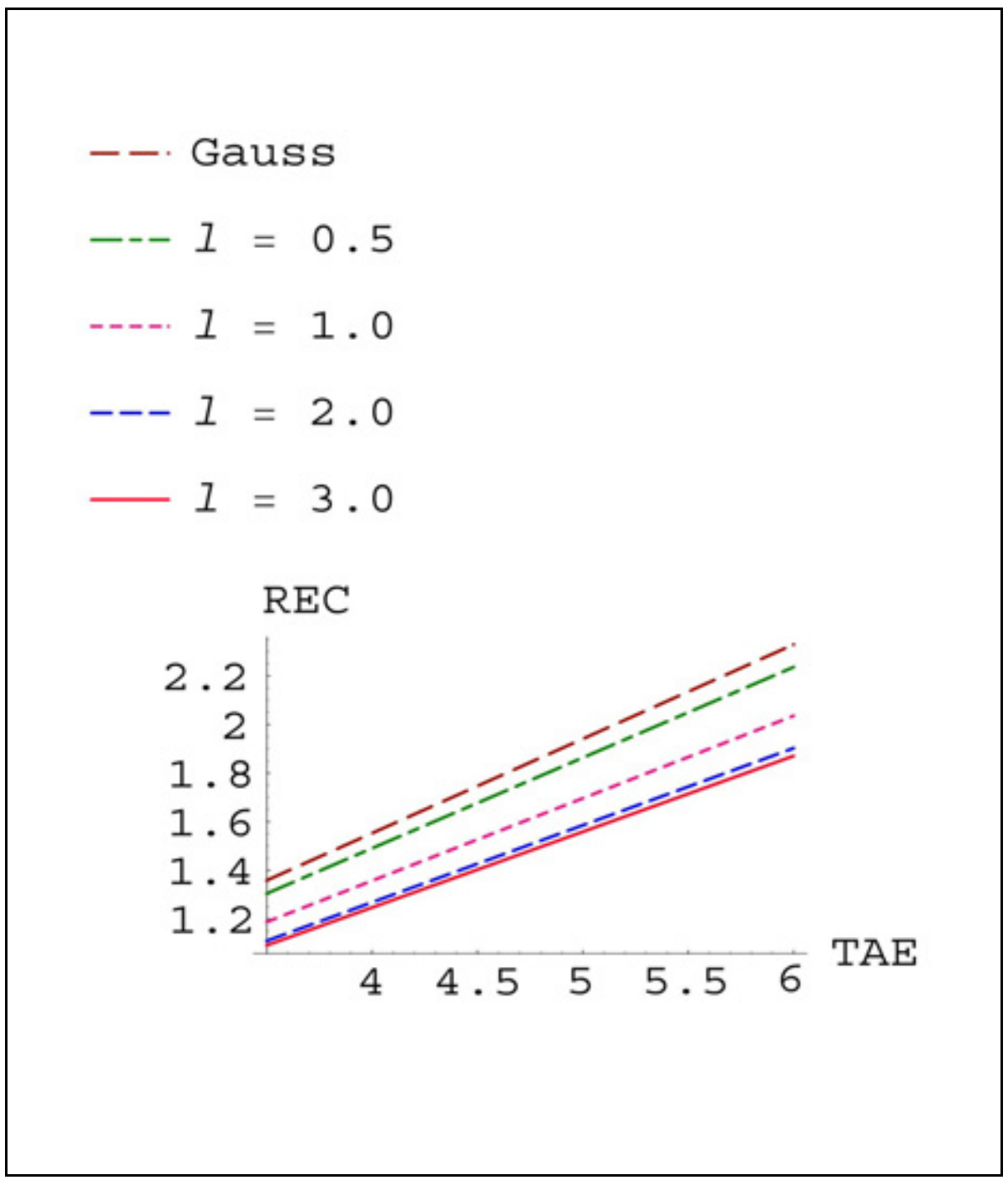

**Figure 6:** The critical random error (REC) versus the total allowable analytical error (TAE), with the Gaussian and the proposed distributions

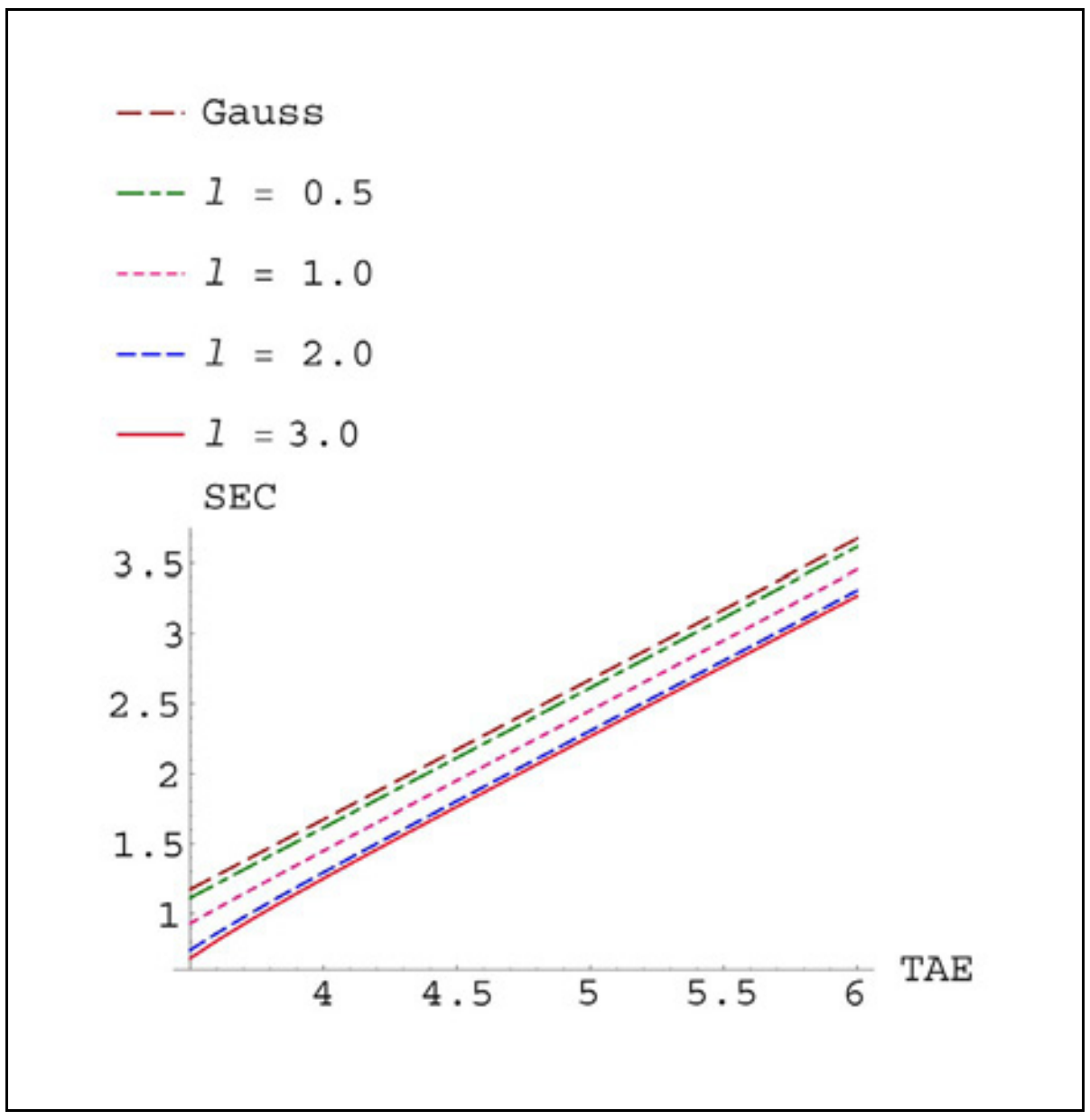

**Figure 7:** The critical systematic error (SEC) versus the total allowable analytical error (TAE), with the Gaussian and the proposed distributions

The Figures 8-9 present the probabilities for critical random and systematic error detection using the *S*(3) QC rule. These probabilities are decreasing with *l*, as well.

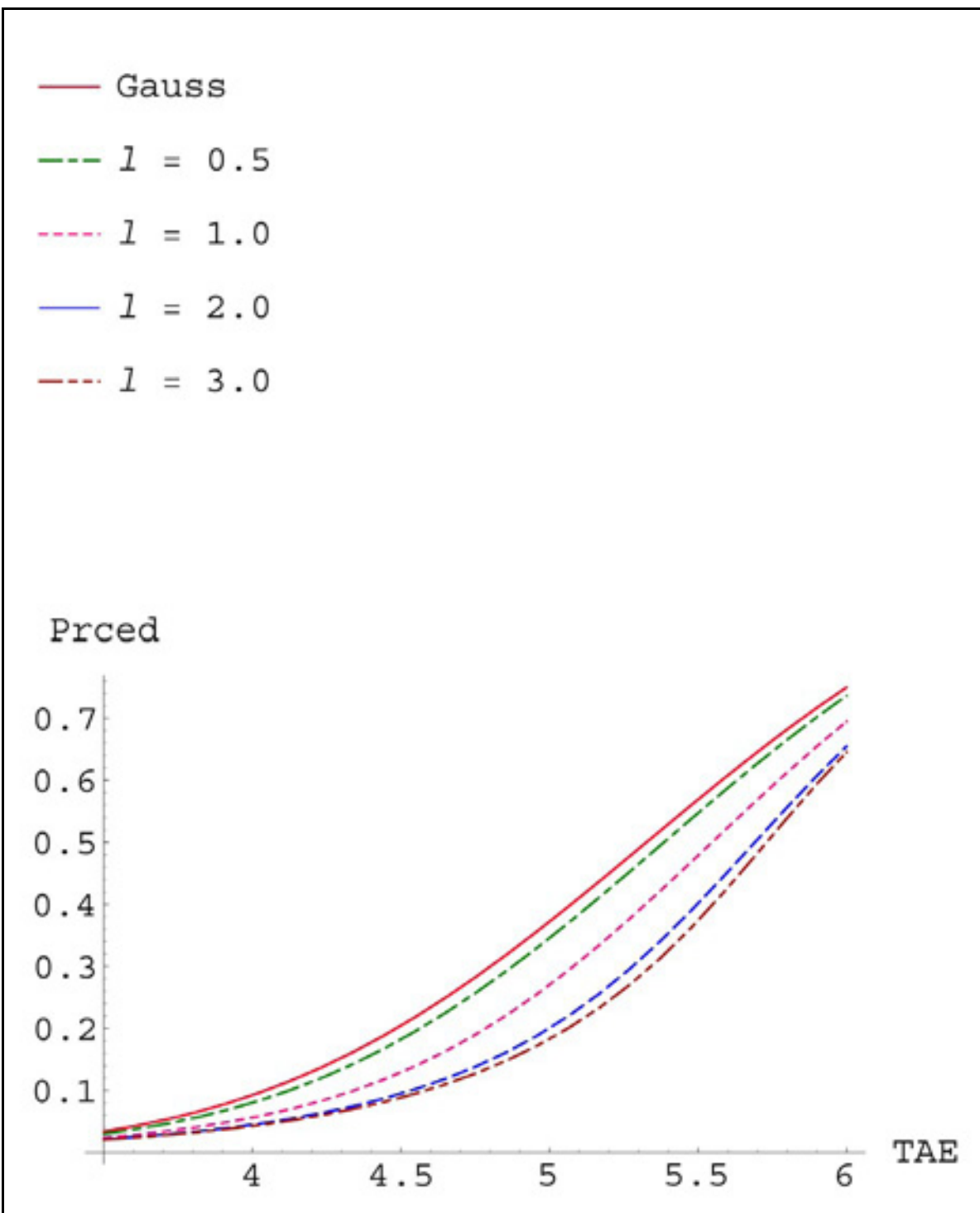

**Figure 8:** The probabilities for critical random error detection (Prced) with the Gaussian and the proposed distributions, using the QC rule *S*(3)

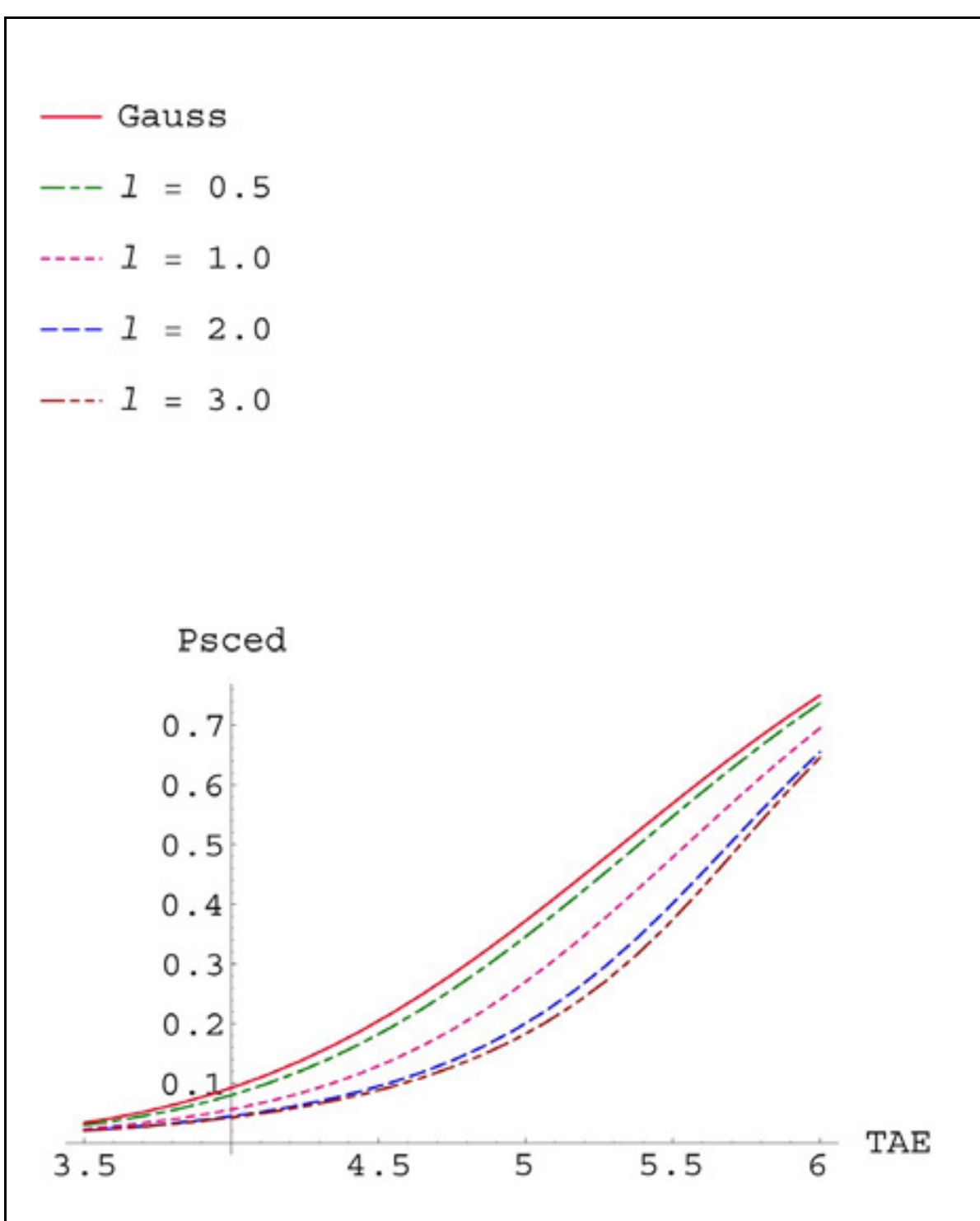

**Figure 9:** The probabilities for critical systematic error (Psced) detection with the Gaussian and the proposed distributions, using the QC rule *S*(3)

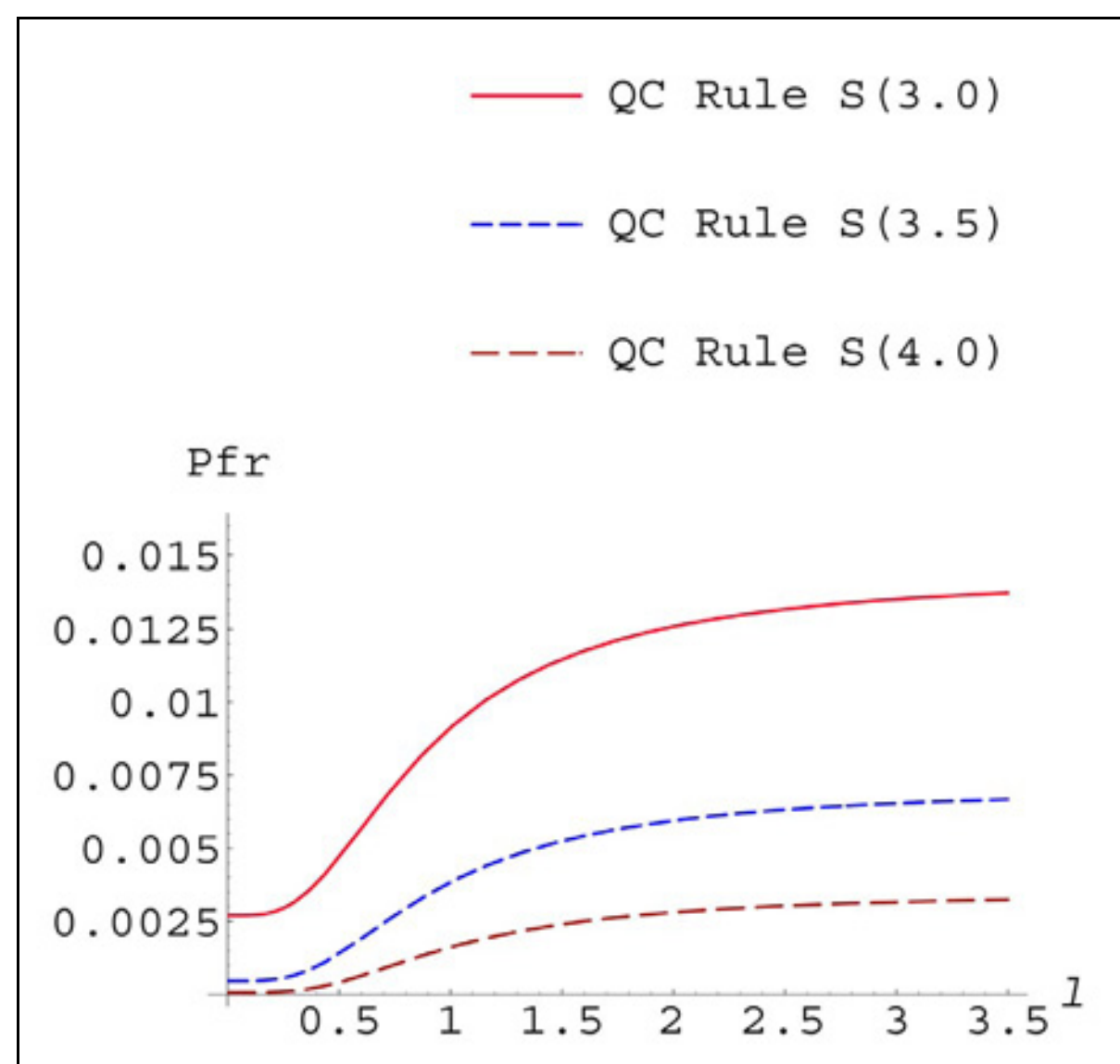

**Figure 10:** The probabilities for false rejection (Pfr) of the QC rules *S*(3.0), *S*(3.5), and *S*(4.0), versus *l*, with the proposed distribution

The Figure 10 presents the plot of the probability for false rejection of the QC rules *S*(3.0), *S*(3.5), and *S*(4.0) versus *l*, with the proposed distribution.

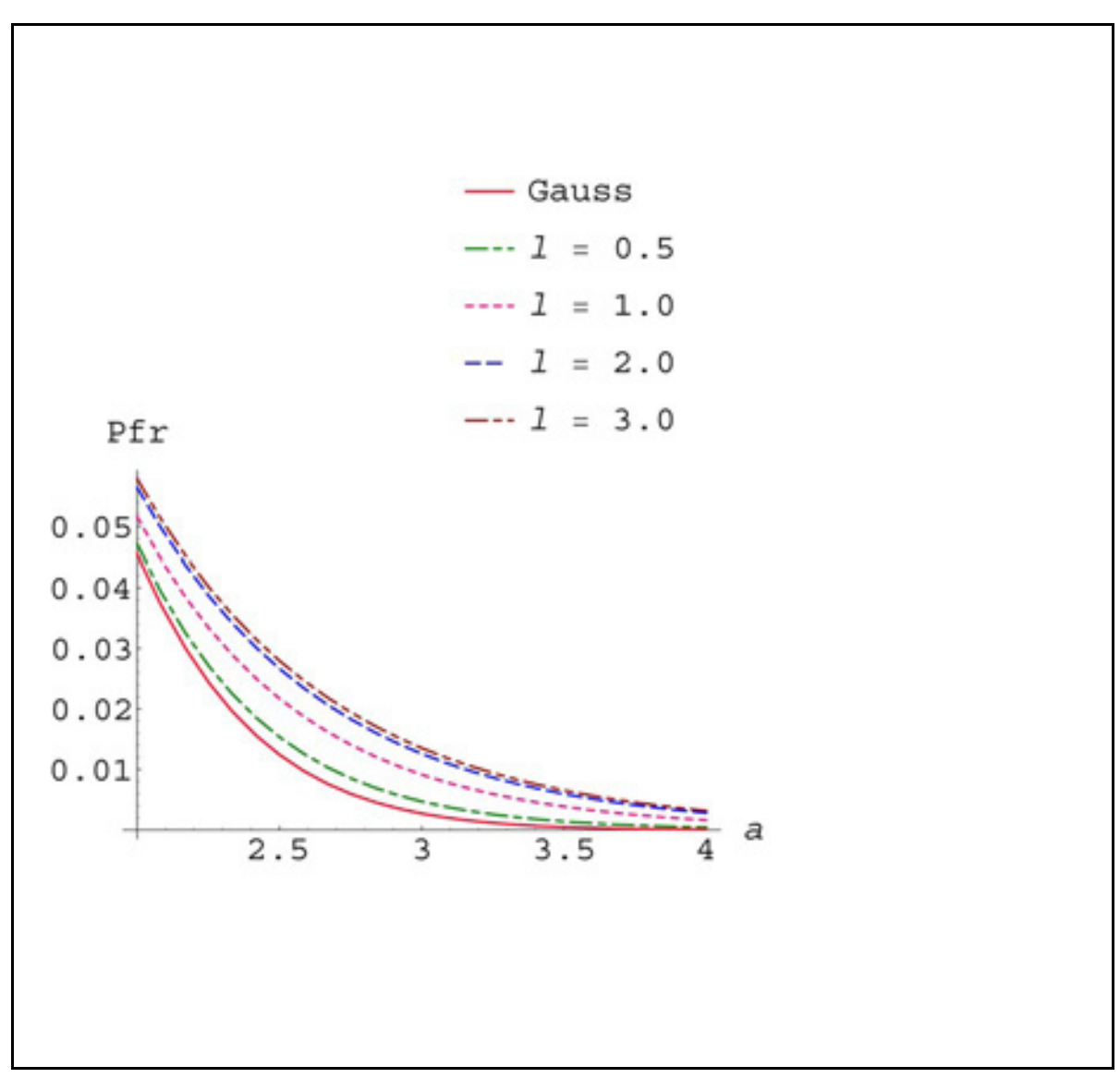

**Figure 11:** The probabilities for false rejection (Pfr) of the QC rules *S*(*a*), versus *a*, with the Gaussian and the proposed distributions

The Figure 11 presents the probabilities for false rejection of the QC rules *S*(*a*), versus *a*, with the Gaussian and the proposed distributions. The probabilities for false rejection are increasing with *l*.

Finally, the figures 12-15 present the probabilities for error detection of the QC rules *S*(3), and *S*(4), with the Gaussian and the proposed distributions.

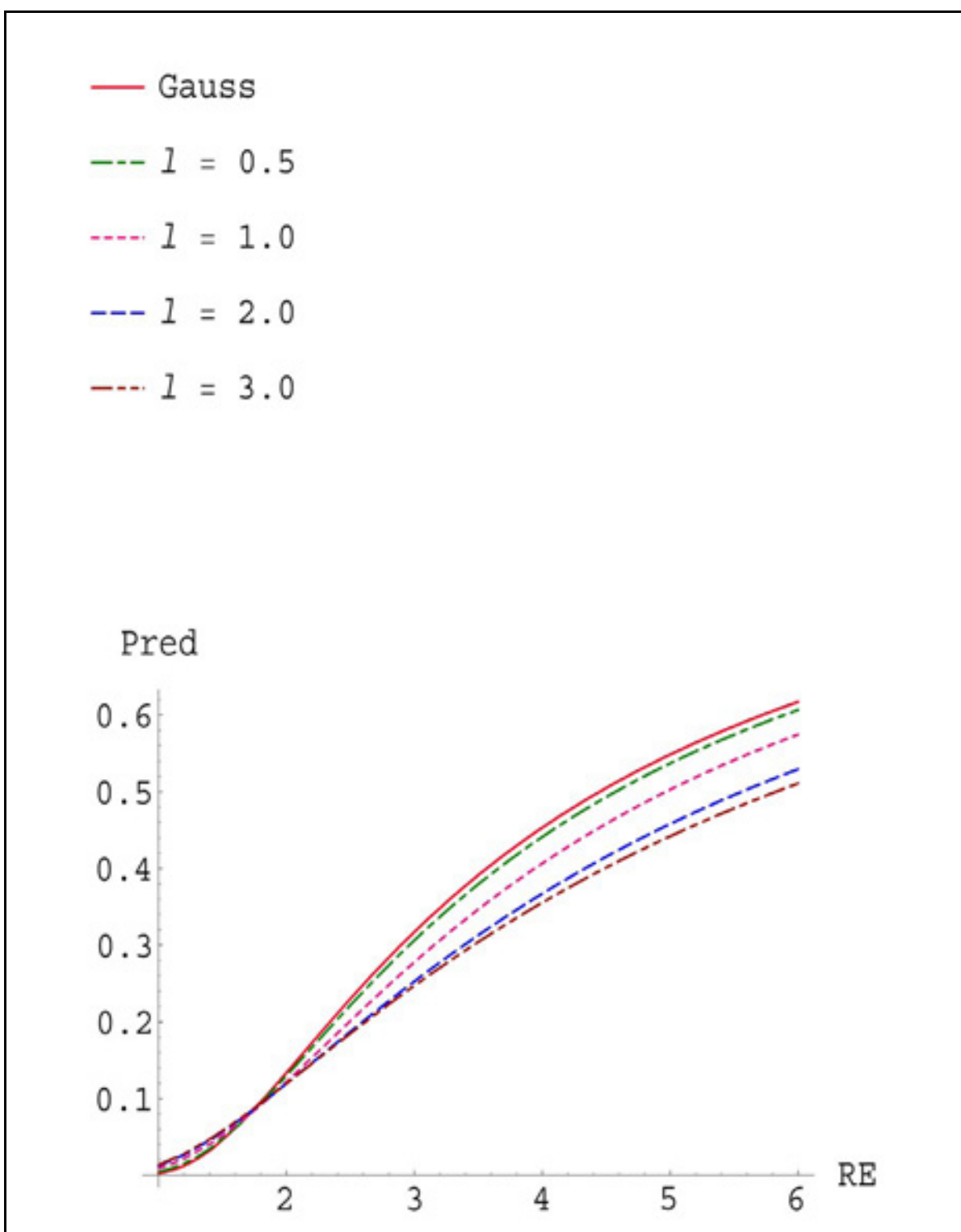

**Figure 12:** The probabilities for random error detection (Pred) of the QC rule *S*(3), with the Gaussian and the proposed distributions

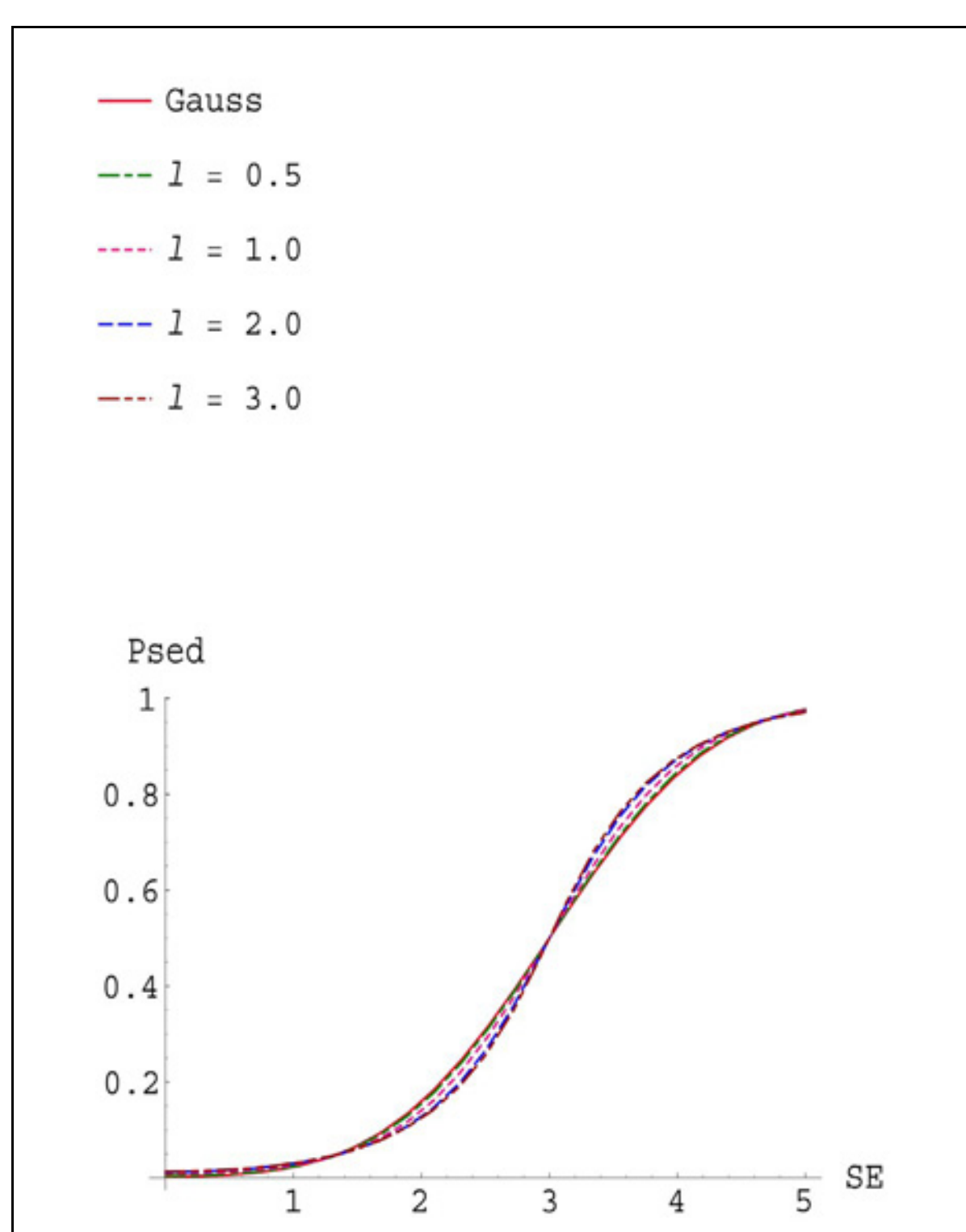

**Figure 13:** The probabilities for systematic error detection (Psed) of the QC rule *S*(3), with the Gaussian and the proposed distributions

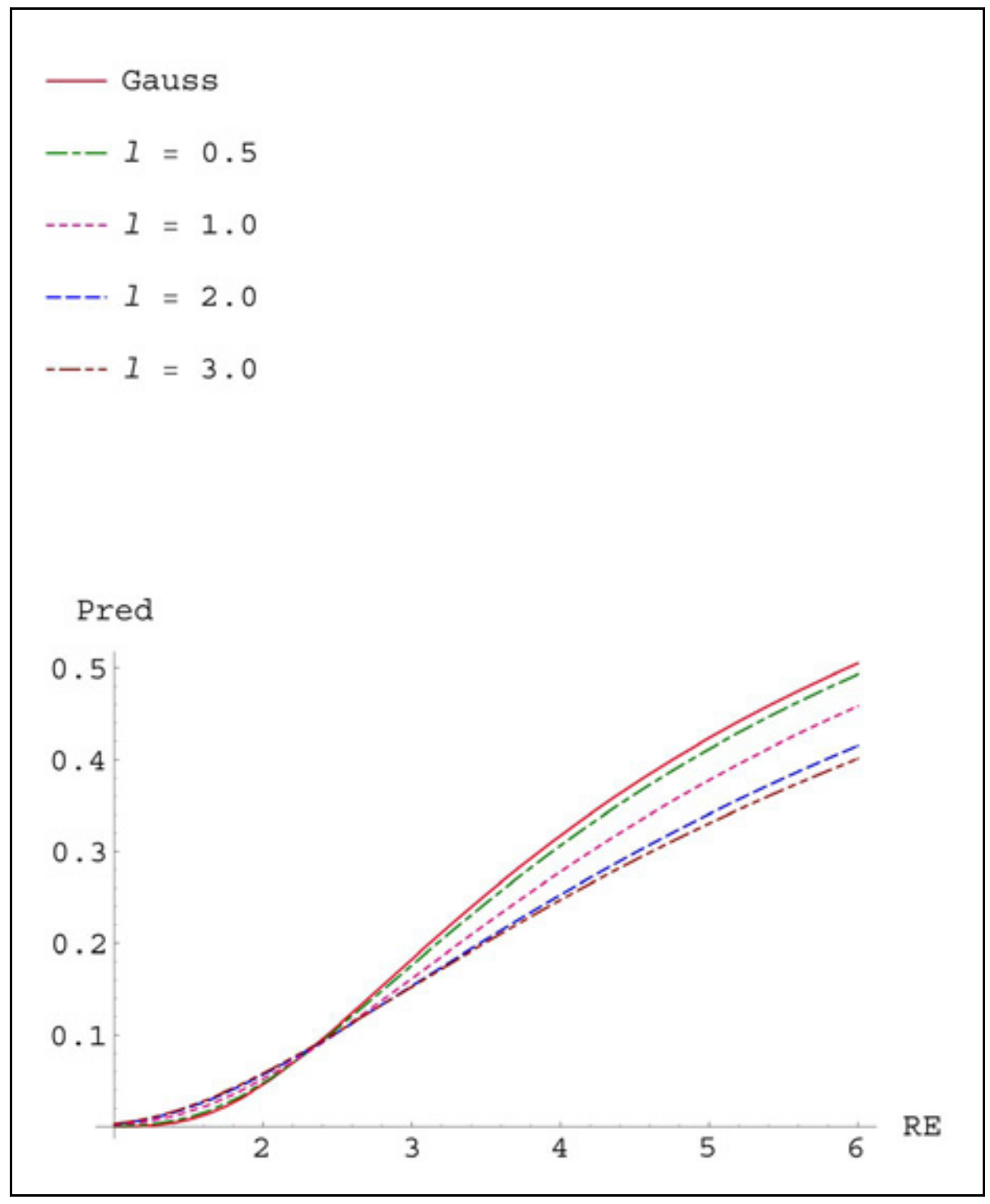

**Figure 14:** The probabilities for random error (Pred) detection of the QC rule *S*(4), with the Gaussian and the proposed distributions

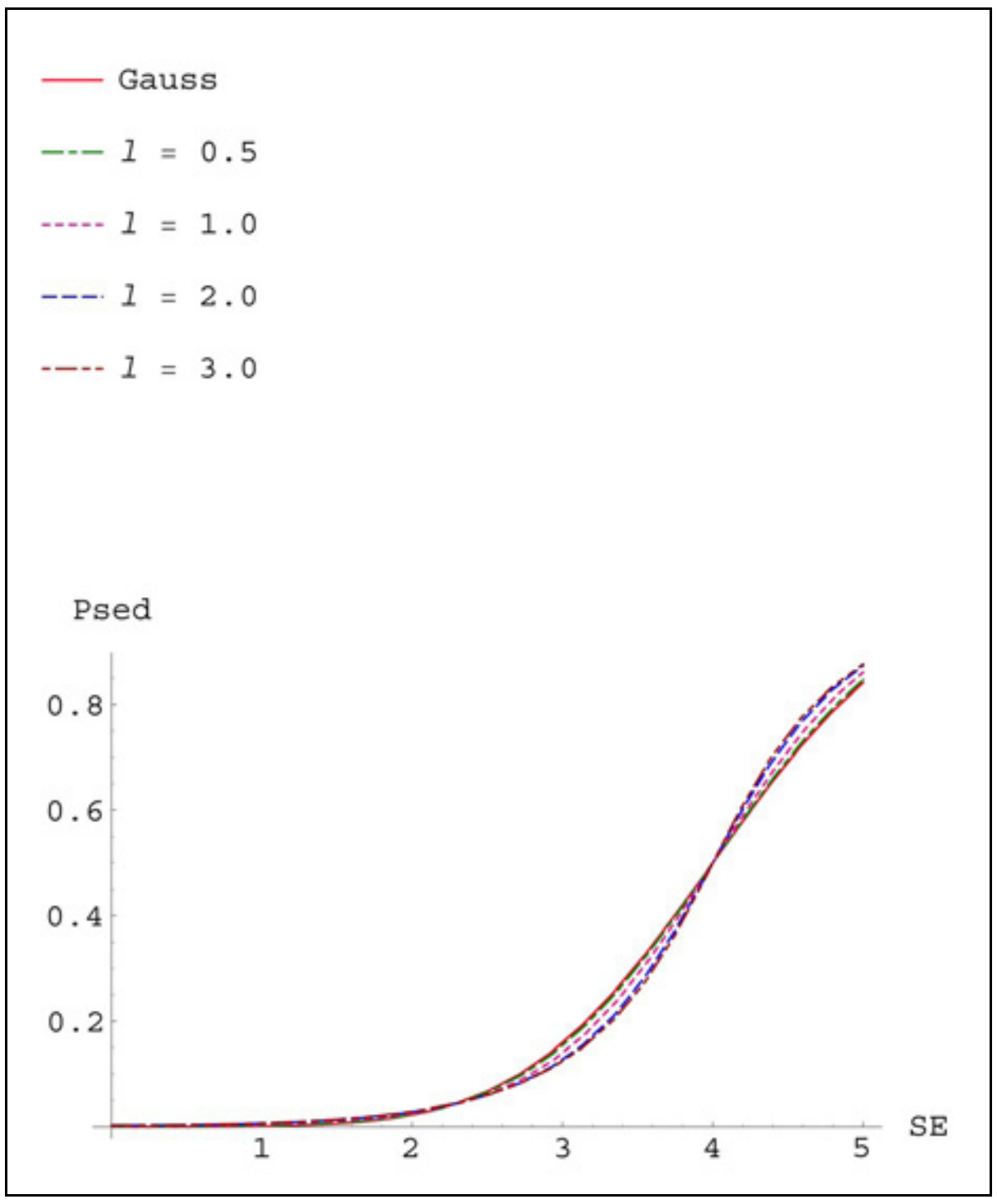

**Figure 15:** The probabilities for systematic error detection (Psed) of the QC rule *S*(4), with the Gaussian and the proposed distributions

## 4. Discussion

Analytical chemistry is usually nonlinear as chemical reactions with strictly first-order kinetics are rare [10, 12]. In addition, the analytical systems in clinical chemistry are becoming increasingly complex, as it is demonstrated by the development of the ultra sensitive assays. As Goldenfeld and Kadanoff have pointed out, improbable events are much more likely with complex nonlinear systems [3]. Their probability can often be approximated by an asymptotically exponential [11] or Laplacian distribution.

The exponential distribution is quite common in physics and physical chemistry. In addition, assuming that $x$ is a random variable distributed exponentially and $h>0$, then the ratio between the probabilities $P(h)/P(x>h)$ is independent of $h$. Consequently, it is scale invariant. Scale invariance is often met in complex nonlinear systems [1] and it is a basic concept of the fractal geometry [8].

However, analytical systems can be approximated by linear models that imply a Gaussian distribution of the analytical error. Nevertheless, linear models fail to explain the experimental finding that in clinical chemistry very large errors are much more probable than it is expected with the Gaussian distribution [13].

Although, some of the large errors are assignable-cause errors, I suggest that there is a nonlinear component of error that is inherent in the analytical processes. In addition I propose that the error of some analytical systems may be approximated by the sum of a linear component of error with Gaussian distribution and a nonlinear component with Laplacian distribution. Consequently, some measurements that are considered as outliers assuming the Gaussian distribution are probable with the proposed distribution.

The estimated probabilities for rejection of the normality hypothesis of simulated series of measurements with the proposed distribution and $l<0.7$, are approximately equal to the respective probabilities of the Gaussian series of measurements (Figure 5), although the probabilities for errors >7SD are up to six orders of magnitude greater than the respective ones of the Gaussian distribution (Table 1). Accordingly, small series of measurements with the proposed distribution could be considered as Gaussian, using some tests of normality.

Assuming the proposed distribution, the critical random and systematic errors can be significantly less than the respective ones of the Gaussian distribution (Figures 6-7). In addition, the probabilities for critical random and systematic error detection are significantly less (Figures 8-9). The probabilities for false rejection are greater than the respective probabilities with the Gaussian distribution (Figures 10-11). The probabilities for error detection differ as well (Figures 12-15).

Therefore, the existence of a nonlinear component of error affects significantly QC. To improve the QC planning process [15] we should explore the existence of a nonlinear component of error, including the so-called "outliers" in the series of the analytical measurements.

The maximum likelihood function can be used to estimate the parameter $l$ of very large series of measurements $x_i$ with the proposed distribution $f(x_i,l)$, but further research is needed:

a) To select useful [12] goodness-of fit-tests for the proposed distribution, possibly based on the empirical distribution functions statistics [9].
b) To explore alternative probability density functions, possibly asymptotically exponential [11], that fit the experimental data.
c) To explore additional nonlinear components of the analytical error, possibly caused by random multiplicative processes [6].
d) To optimize the QC planning process assuming a nonlinear component of the analytical error, with numerical methods as well as with powerful heuristic tools as the genetic algorithms [4, 5].

## 5. Conclusion

In conclusion, the experimentally found probabilities of very large analytical errors imply the existence of a nonlinear component of the analytical error that affects QC.

## 6. Acknowledgements

I wish to thank Theophanis T. Hatjimihail of the Hellenic Complex Systems Laboratory, for his continuing support and encouragement, as well as for his specific suggestions.

## 8. Abbreviations

The following non standard abbreviations are used in the figures:
p: Probability
SD: Standard Deviation
r: Ratio
REC: Critical Random Error (in SD units)
SEC: Critical Systematic Error (in SD units)
TAE: Total Allowable Analytical Error (in SD units)
Prced: Probability for Random Critical Error Detection
Psced: Probability for Systematic Critical Error Detection
Pfr: Probability for False Rejection
Pred: Probability for Random Error Detection
RE: Random Error (in SD units)
Psed: Probability for Systematic Error Detection
SE: Systematic Error (in SD units)

## 9. References

1. Badii R, Politi A. Complexity: hierarchical structures and scaling in physics. Cambridge: Cambridge University Press, 1997.
2. Duncan AJ. Tests of normality. Duncan AJ. Quality Control and Industrial Statistics. Homewood, Illinois: Irwin, 1986: 634-45.
3. Goldenfeld N, Kadanoff LP. Simple lessons from complexity. Science 1999; 284(5411):87-9.
4. Hatjimihail AT. Genetic algorithms-based design and optimization of statistical quality- control procedures [published erratum appears in Clin Chem 1994 Feb;40(2):267]. Clin Chem 1993; 39(9):1972-8.

5. Hatjimihail AT, Hatjimihail TT. Design of statistical quality control procedures using genetic algorithms. In: Eshelman Larry J., Editor. Proceedings of the Sixth International Conference on Genetic Algorithms. San Francisco, California: Morgan Kauffman, 1995: 551-7.
6. Kadanoff LP. Statistical physics: statistics, dynamics and renormalization. Singapore: World Scientific, 2000: 84-91.
7. Linnet K. Choosing quality-control systems to detect maximum clinically allowable analytical errors. Clin Chem 1989; 35(2):284-8.
8. Mandelbrot BB. Fractals and scaling in finance: discontinuity, concentration, risk. New York: Springer-Verlag, 1997.
9. Puig P, Stephens MA. Tests of fit for the Laplace distribution, with applications. Technometrics 2000; 42(4):417-24.
10. Scott SK. Chemical chaos. New York: Oxford University Press, 1991.
11. Shraiman BI, Siggia ED. Lagrangian path integrals and fluctuations in random flow. Physical Review E 1994; 49(4):2912-27.
12. Whitesides GM, Ismagilov RF. Complexity in chemistry. Science 1999; 284(5411):89-92.
13. Witte DL, VanNess SA, Angstadt DS, Pennell BJ. Errors, mistakes, blunders, outliers, or unacceptable results: how many? Clin Chem 1997; 43(8 Pt 1):1352-6.
14. Wolfram S. The Mathematica book. 4th edition. Champaign, Illinois; Cambridge: Wolfram Media /Cambridge University Press, 1999.
15. Woodall WH. Controversies and contradictions in statistical process control. Journal of Quality Technology 2000; 32(4):341-50.